%% file: iclr2025_conference.tex
\documentclass{article} % For LaTeX2e
\usepackage{iclr2025_conference,times}

\input{math_commands.tex}

\usepackage{hyperref}
\usepackage{url}

\usepackage{emoji}
\usepackage{booktabs}
\usepackage{tabularx}
\usepackage{array}
\usepackage{multirow}
\usepackage{makecell}
\usepackage{xcolor}
\usepackage{subcaption}
\usepackage{colortbl}
\usepackage{adjustbox} % In your preamble
\usepackage{array}     % For m{} column type
\usepackage{ulem}

\usepackage{enumitem}

\newcommand{\celllines}[1]{%
  \begin{tabular}[t]{@{}l@{}}#1\end{tabular}%
}

\newcommand{\nodotc}[1]{%
  \begin{minipage}[t]{\linewidth}
    \begin{itemize}[label={},labelsep=0pt,leftmargin=0pt,
                    topsep=0pt,itemsep=0pt,parsep=0pt,partopsep=0pt]
      #1
    </begin{itemize}
  \end{minipage}%
}

\title{Position: Privacy is not just memorization!}

\author{\textbf{Niloofar Mireshghallah}\textsuperscript{1}\thanks{Equal Contribution, order decided by coin toss} \quad \textbf{Tianshi Li}\textsuperscript{2}\textsuperscript{$*$} \\
\textsuperscript{1}Carnegie Mellon University \quad \textsuperscript{2}Northeastern University \\
\texttt{niloofar@cmu.edu} \quad \texttt{tia.li@northeastern.edu}
}

\iclrfinalcopy % Uncomment for camera-ready version, but NOT for submission.
\begin{document}

\maketitle

\begin{abstract}
The discourse on privacy risks in Large Language Models (LLMs) has disproportionately focused on verbatim memorization of training data, while a constellation of more immediate and scalable privacy threats remain underexplored. \textit{This position paper argues that the privacy landscape of LLM systems extends far beyond training data extraction, encompassing risks from data collection practices, inference-time context leakage, autonomous agent capabilities, and the democratization of surveillance through deep inference attacks}. We present a comprehensive taxonomy of privacy risks across the LLM lifecycle---from data collection through deployment---and demonstrate through case studies how current privacy frameworks fail to address these multifaceted threats. Through a longitudinal analysis of 1,322 AI/ML privacy papers published at leading conferences over the past decade (2016--2025), we reveal that while memorization receives outsized attention in technical research, the most pressing privacy harms lie elsewhere, where current technical approaches offer little traction and viable paths forward remain unclear. We call for a fundamental shift in how the research community approaches LLM privacy, moving beyond the narrow focus of current technical solutions and embracing interdisciplinary approaches that address the sociotechnical nature of these emerging threats.

\end{abstract}

\section{Introduction}

Large Language Models are fundamentally data-driven systems, trained on vast corpora scraped from the web \citep{brown2020language}, user interactions \citep{ouyang2022training}, and increasingly, real-time retrieval systems \citep{lewis2020retrieval}. While privacy concerns have rightfully emerged as these models consume unprecedented amounts of personal data, the research community's response has been disproportionately narrow---fixating almost exclusively on verbatim memorization and training data extraction \citep{carlini2021extracting,carlini2023quantifying}. In this paper, we advance a critical position: \textit{Privacy in LLM systems is not just about memorization. It encompasses how providers extract consent through deceptive interfaces, how autonomous agents exfiltrate user data without regard for privacy norms, how systems aggregate scattered information to reveal intimate details and provide answers to secondary questions used for password recovery, and how models can transform innocuous public data into targeted surveillance or stalking capabilities.}

We systematically categorize the privacy landscape by first identifying \textbf{three types of data} flowing through LLM ecosystems: \textit{(i) user interaction data} encompassing prompts, feedback, and conversation histories; \textit{(ii) system-retrieved data} from RAG pipelines, APIs, and real-time sources; and \textit{(iii) publicly available data} including web corpora with embedded credentials and personal information (Section~\ref{sec:data_effect}). These data types interact to create \textbf{five distinct categories of privacy incidents} (Table~\ref{tab:taxonomy}): beyond \textit{training data leakage via regurgitation}, we identify \textit{direct chat leakage} through provider breaches and deceptive policies, \textit{indirect context leakage} via autonomous agents and prompt injection, \textit{indirect attribute inference} where LLMs deduce sensitive information from innocuous inputs, and \textit{direct attribute aggregation} that weaponizes dispersed online information (Section~\ref{sec:regurg}--\ref{sec:teles}). Each incident type presents unique threats---from agents exfiltrating database contents through compromised RAG systems to LLMs inferring precise locations from seemingly anonymous photos---fundamentally transforming these systems from passive data stores into active, privacy-violating inference engines. \textit{The harms extend beyond privacy into security domains}, as information aggregated by deep research agents can be exploited to answer seemingly innocuous questions---such as ``What’s Alice’s pet cat’s name?'' (see a real example in Figure~\ref{fig:pet_cat})---which in turn can enable secondary attacks like password retrieval and account theft~\citep{little2024secure}.

Having identified five distinct categories of privacy incidents in LLM systems (Section~\ref{sec:exposure})---from training data leakage to inference attacks and aggregation threats---a critical question emerges: does the research community's focus align with these real-world privacy risks? To answer this, we conduct a systematic analysis of AI/ML privacy research published at leading conferences over the past decade (2016--2025). Our findings (Section~\ref{sec:exp}) reveal a striking misalignment between research priorities and practical privacy threats. While 92\% of papers focus on training data memorization and cryptographic protections against direct chat leakage, the remaining incident types—indirect attribute inference, agent-based context leakage, and direct attribute aggregation—collectively receive less than 8\% of research attention suggesting disciplinary blind spots that leave critical vulnerabilities unaddressed.

This paper paves a path forward through technical interventions that work today (local data minimization, hybrid architectures, privacy-aligned post-training), sociotechnical approaches that empower users (contextual integrity frameworks, awareness tools, tradeoff visualization), and policy reforms that address power asymmetries. We demonstrate that privacy protection requires moving beyond the narrow lens of memorization to address deceptive consent, inference attacks, and the commodification of conversation. The privacy challenges are sociotechnical, not purely algorithmic—requiring collaboration between technologists, designers, policymakers, and affected communities. The rest of the paper is organized as follows:
\input{tables/table_overview}

\clearpage

\noindent\textbf{\S2 What Data is Affected?}
\begin{itemize}[leftmargin=1.5em, topsep=0pt, itemsep=2pt]
  \item \S2.1 Data Collection and Retention Policies
  \begin{itemize}[leftmargin=1.5em, topsep=0pt, itemsep=0pt]
    \item \S2.1.1 What is Explicit Consent? The Default Opt-in Setting
    \item \S2.1.2 Do Users Really Have a Choice? Opt-out and Other Limitations
  \end{itemize}
  \item \S2.2 Different Types of Data in the LLM Ecosystem
  \begin{itemize}[leftmargin=1.5em, topsep=0pt, itemsep=0pt]
    \item \S2.2.1 User Interaction Data
    \item \S2.2.2 System Retrieved Data
    \item \S2.2.3 Publicly Available Data
  \end{itemize}
\end{itemize}

\noindent\textbf{\S3 How is the Data Being Exposed?}
\begin{itemize}[leftmargin=1.5em, topsep=0pt, itemsep=2pt]
  \item \S3.1 Training Data Leakage via Regurgitation
  \item \S3.2 Direct Chat Leakage via Uninformed Consent or Compromised Provider
  \item \S3.3 Indirect Chat and Context Leakage via Input-Output Flow
  \item \S3.4 Privacy Under the Microscope: Indirect Attribute Inference
  \item \S3.5 Privacy Through the Telescope: Direct Attribute Aggregation
\end{itemize}

\noindent\textbf{\S4 A Decade of AI/ML Privacy Research: Trends from Leading ML, NLP, and S\&P Conferences}
\begin{itemize}[leftmargin=1.5em, topsep=0pt, itemsep=2pt]
  \item \S4.1 Corpus
  \item \S4.2 Annotation Pipeline
  \item \S4.3 Results
\end{itemize}

\noindent\textbf{\S5 Technical Solutions and Beyond: A Roadmap Forward}
\begin{itemize}[leftmargin=1.5em, topsep=0pt, itemsep=2pt]
  \item \S5.1 Technical Interventions
  \item \S5.2 Sociotechnical Approaches
  \item \S5.3 Policy and Governance
\end{itemize}

\noindent\textbf{\S6 Conclusion}

\section{What Data is Affected?}\label{sec:data_effect}

The scope of data at risk in LLM systems extends far beyond training corpora. To understand the privacy risks posed by LLMs beyond verbatim memorization and regurgitation of often publicly-available pretraining data~\citep{carlini2021extracting,carlini2023quantifying}, we must expand our view from the model in isolation to the entire LLM ecosystem. This ecosystem encompasses data collection and curation, model training, deployment infrastructure, serving systems, and third-party API wrappers~\citep{wang2025large,bommasani2021opportunities}. Recent research has identified systematic vulnerabilities across this ecosystem, from poisoned RAG systems~\citep{zou2025poisonedrag} to insecure third-party app stores~\citep{hou2025insecurity}. Each component in this ecosystem touches different types of data, creating compounding privacy vulnerabilities that current research has only begun to explore~\citep{siyan2024papillon,liu2025protecting}.

We categorize the data affected by the LLM ecosystem into three distinct types: \textit{user interaction data}, \textit{system-retrieved data}, and \textit{publicly available data}. We define each category and analyze the unique privacy risks they present. These categories are not mutually exclusive—their risks compound when data flows between them, as we demonstrate through our incident taxonomy in Table~\ref{tab:taxonomy}. Understanding these data types and their interconnections is crucial for developing comprehensive privacy protections that address the full scope of LLM-related risks, moving beyond the narrow focus on training data extraction that has dominated prior work.

\begin{figure}
    \centering
    \includegraphics[width=1\linewidth]{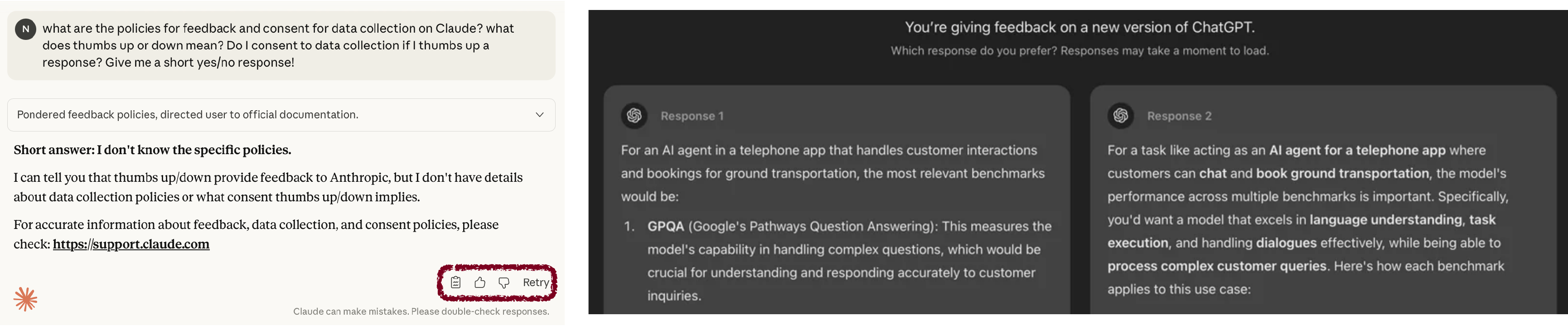}
    \caption{Examples of `automatic' consent mechanisms deployed by Anthropic (giving a thumbs up or down on Claude responses opts the conversation into data collection, left) and OpenAI (selecting a response records the conversation in ChatGPT, right).}
    \label{fig:consent}
\end{figure}

\subsection{Data Collection and Retention Policies}
Before we discuss the different types of data impacted by the LLM ecosystem, we need to examine how LLM providers define consent, implement opt-out mechanisms, and retain user data in practice.

\subsubsection{What is Explicit Consent? The Default Opt-in Setting}
In the context of LLM services, explicit consent has been fundamentally redefined in ways that would be unrecognizable under traditional privacy frameworks. Let's have a look at the different policies set by prominent frontier LLM providers:

\begin{itemize}
    \item \textbf{Anthropic} \sout{employed what appeared to be the most restrictive approach on paper, stating they ``will not use your Inputs or Outputs to train our generative models unless you've explicitly reported the materials to us}. This opt-in model applied equally to free and paid individual users, though enterprise customers received additional protections through Zero Data Retention agreements. \textbf{This restrictive approach was abandoned in September 2025 (the time of writing)}. In a reversal from their privacy-first stance, Anthropic now requires users to explicitly opt-out by September 28, 2025, or their conversations and coding sessions will be used to train AI models. The new policy extends data retention from 30 days to 5 years for users who do not opt out~\cite{anthropic_privacy_update_2025}. This shift affects all consumer tiers (Claude Free, Pro, and Max), though enterprise customers under Commercial Terms maintain their privacy protections. \textbf{Note that automatic consent still occurs when users provide feedback through thumbs up/down mechanisms (see \autoref{fig:consent}), regardless of their opt-out status~\cite{anthropic_privacy_2025}.}

    \item \textbf{OpenAI} defines consent more broadly, with free users' data used for training by default unless they actively opt out. Their policy states that ``Content'' includes ``any data, files, or information you provide through our Services,'' encompassing prompts, uploads, and all interactions~\cite{openai_data_faq}. Paid tiers like ChatGPT Plus follow the same default training usage as free users, while only enterprise customers receive automatic opt-out protections~\cite{openai_enterprise}.
    
    \item \textbf{Google Gemini} takes the most expansive approach, with ``Gemini Apps Activity on by default if you are 18 or older,'' automatically collecting ``your chats, what you share with Gemini Apps (like files, images, screens), related product usage information, your feedback, and location info''~\citep{gemini_privacy_hub}. While Google One AI Premium subscribers receive some enhanced protections where ``Google doesn't use your prompts or responses to improve our products,'' feedback and certain metadata remain subject to collection.
    
    \item \textbf{Grok/xAI} implements the most aggressive collection, with all X users automatically opted-in to data sharing for AI training. The November 2024 policy update expanded this to include sharing with ``third-party collaborators'' beyond just xAI~\citep{techcrunch_grok_2024}. This includes all public posts, interactions, voice inputs, and cross-platform integration data when using X credentials~\citep{xai_privacy}.
\end{itemize}

\begin{figure}
    \centering
    \includegraphics[width=1\linewidth]{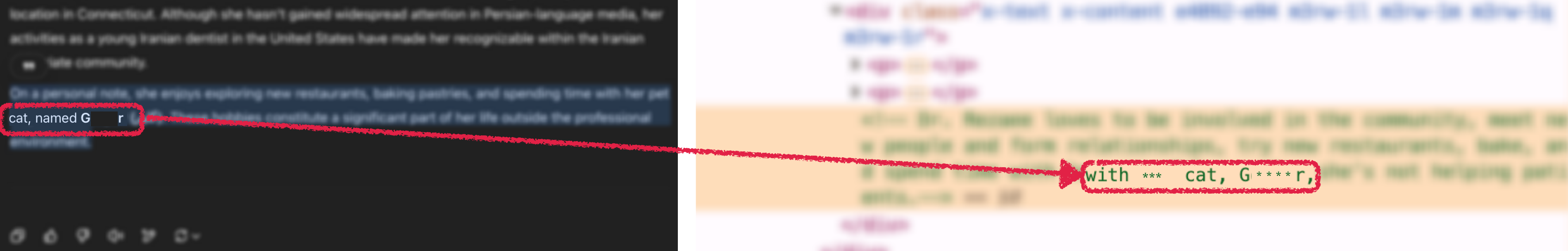}
    \caption{Example of a redacted query to ChatGPT’s deep research: It uncovers the name of an individual’s pet cat from a comment embedded in an HTML tag. This is particularly concerning, as such niche information is often used in password recovery, which could facilitate account theft and create security risks~\citep{little2024secure}.}
    \label{fig:pet_cat}
\end{figure}

In summary all major providers now operate on opt-out models that favor data collection. While the majority offer paid tiers that ostensibly provide enhanced privacy protections, even paid subscriptions contain hidden vulnerabilities where various mechanisms automatically opt you in or grant consent on your behalf. Let's examine these practices.

\paragraph{Thumbs up or down? You just consented to 10 years of data retention, even as a paid user!}
Perhaps most concerning is how all providers exploit feedback mechanisms to bypass privacy protections. OpenAI explicitly states: ``\textit{Even if you've opted out of training, if you choose to provide feedback (for instance, by selecting thumbs up or thumbs down), the entire conversation associated with that feedback may be used to train our models}''~\citep{openai_feedback_2024}. This creates a particularly deceptive practice where a simple evaluative gesture grants comprehensive training rights that override all other privacy settings. The ``which is better'' comparison interfaces employed by these services similarly trigger automatic data collection rights.

Anthropic retains feedback-related conversations ``\textit{in our secured back-end for up to 10 years},'' while Google keeps such data ``for up to 3 years, disconnected from your Google Account'' but \textit{immune to user deletion requests}~\citep{anthropic_retention, gemini_privacy_hub}. Even Grok's feedback system creates permanent training data that \textit{cannot be removed from models once processed}~\citep{xai_faq}. \textbf{Note that even after Anthropic's September 2025 policy change requiring explicit opt-out for training, the feedback mechanism still triggers extended retention periods that override user preferences.}

\paragraph{Arbitrary security classifiers can mark to keep your data forever.}
All major providers maintain broad security exceptions that override deletion policies. OpenAI reserves the right to retain data for ``legal or security reasons'' with automated classifiers for abuse detection triggering minimum 30-day retention that \textit{can extend indefinitely}~\citep{openai_privacy_policy}. Anthropic's Constitutional AI classifiers trigger extended retention: ``We retain inputs and outputs for up to 2 years and trust and safety classification scores for \textit{up to 7 years} if you submit a prompt that is flagged by our trust and safety classifiers''~\citep{anthropic_retention}. These classifiers monitor for CBRN content, violence, and other policy violations with \textit{opaque criteria}. Google maintains that ``conversations that have been reviewed or annotated by human reviewers are not deleted when you delete your Gemini Apps activity,'' with 3-year retention for flagged content~\citep{gemini_privacy_hub}. The company's cross-service integration enables data sharing for ``detecting, preventing, and responding to fraud, abuse, security risks, and technical issues'' across all Google properties~\citep{google_information_requests}.

\paragraph{Your conversations persist for years.}
Standard retention periods range from \sout{30 days (Anthropic's default deletion for non-flagged conversations)} \textbf{30 days only for Anthropic users who explicitly opt out (as of September 2025), to 5 years for those who don't opt out}, to 18 months (Google's default Gemini activity retention), with \textit{feedback data retained for 3-10 years regardless of account deletion}~\citep{anthropic_retention, gemini_privacy_hub}. More critically, \textit{a federal court order since May 2025 requires OpenAI to preserve consumer ChatGPT and API customer data indefinitely indefinitely, even if they are deleted by the user}, affecting all consumer users~\citep{nyt_lawsuit_2024, venturebeat2025retention}. 

Training usage depends on both user tier and interaction type: \sout{Anthropic maintains its no-training policy without explicit consent,} \textbf{Anthropic now uses consumer data by default unless users opt out (as of September 2025)}, OpenAI uses free/Plus user data unless opted out while protecting enterprise data by default, Google trains on all free user data while excluding paid user prompts but continuing to use their feedback, and Grok trains on all user data by default with \textit{retroactive model training that cannot be reversed}~\citep{anthropic_privacy_2025, openai_data_faq, gemini_api_terms, silicon_republic_grok}.

\begin{figure}[t]
    \centering
    \begin{subfigure}{0.5\linewidth}
        \centering
        \includegraphics[width=\linewidth]{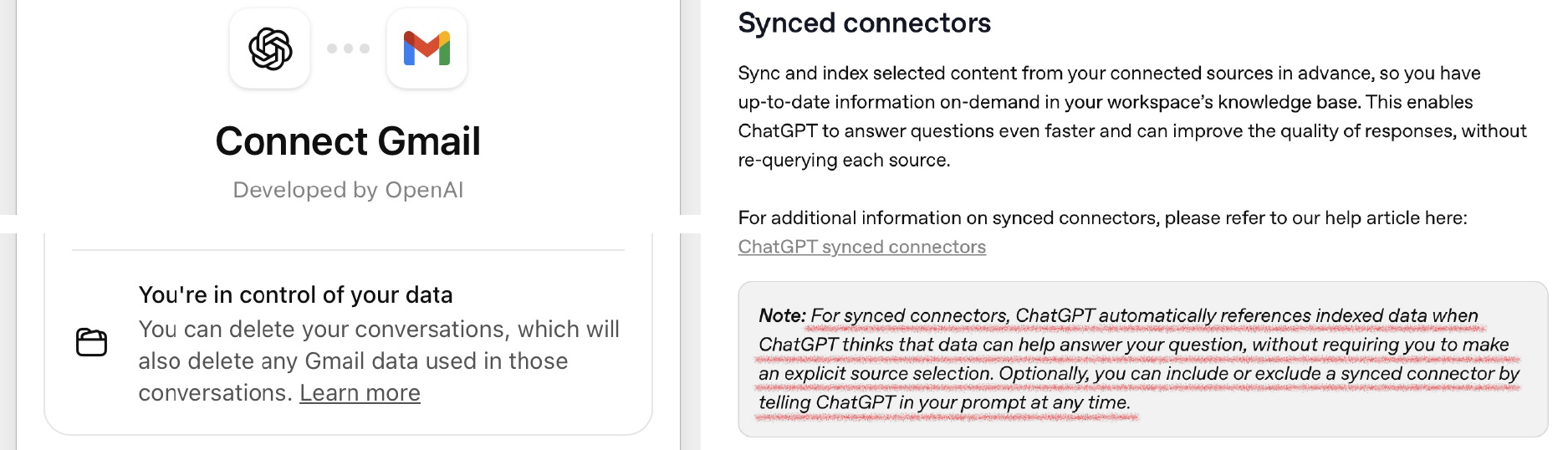}
        \caption{OpenAI Connectors}
        \label{fig:openai-connectors}
    \end{subfigure}
    \hfill
    \begin{subfigure}{0.45\linewidth}
        \centering
        \includegraphics[width=\linewidth]{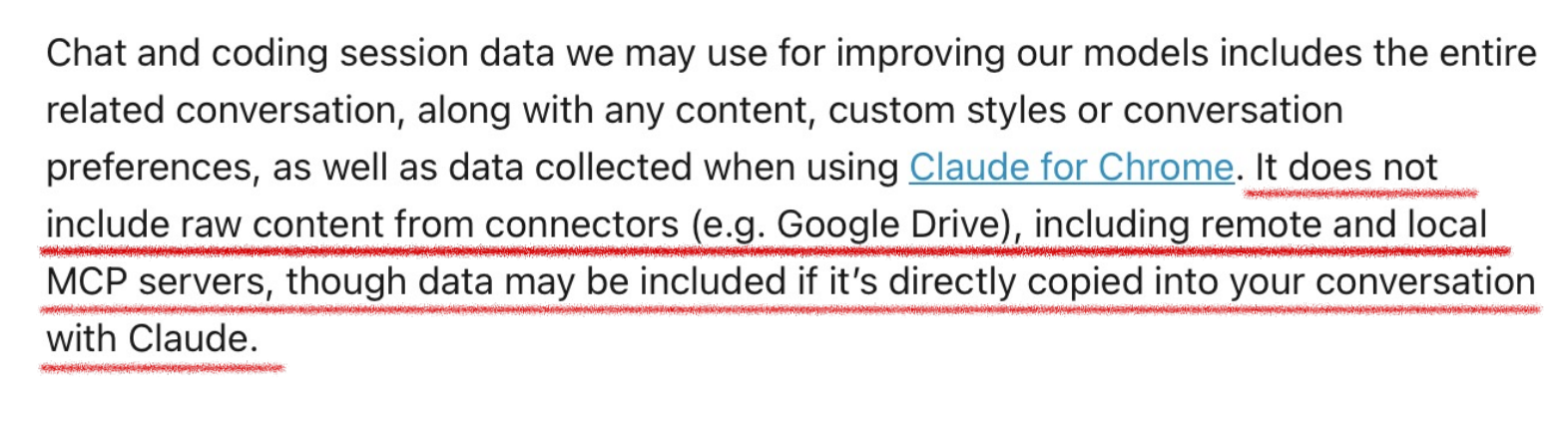}
        \caption{Claude Connectors}
        \label{fig:claude-connectors}
    \end{subfigure}
    \caption{OpenAI and Claude both provide connectors for \textit{automatic} integration of external user data, but the data is often scattered and requires manual deletion to be fully removed.}
    \label{fig:connectors}
\end{figure}
\subsubsection{Do Users Really Have a Choice? Opt-out and Other Limitations}

While providers offer opt-out mechanisms, these systems are deliberately complex and often ineffective, creating barriers that discourage users from exercising privacy rights.

\begin{itemize}
    \item \textbf{Anthropic} provides the most limited options with \textit{no global opt-out for feedback data once submitted} and no retroactive control over previously shared conversations~\citep{anthropic_privacy_2025}. API customers face additional restrictions: ``\textit{For paid API customers, we do not support ad hoc deletion}''~\citep{anthropic_api_deletion}.
    \item \textbf{OpenAI}'s multi-pathway opt-out system—accessible through ChatGPT Settings, privacy.openai.com, or Temporary Chat mode—appears comprehensive but contains critical gaps. The chat history used to be unnecessarily bundled with training controls~\citep{zhang2024fair} until it was separated into distinct controls in April 2024~\citep{bgr_privacy_2024}. Moreover, the ongoing New York Times lawsuit has resulted in \textit{a court order requiring OpenAI to preserve all user data indefinitely}, fundamentally undermining any opt-out preferences~\citep{nyt_lawsuit_2024, magai_court_2025}.
    \item \textbf{Google Gemini}'s opt-out mechanism has drawn particular criticism for its complexity. Users must navigate to myactivity.google.com, locate Gemini-specific controls among dozens of Google services, and even then face limitations: ``Conversations that have been reviewed or annotated by human reviewers are not deleted when you delete your Gemini Apps activity''~\citep{gemini_privacy_hub}. The July 2025 update that \textit{automatically enabled Gemini access to Phone, Messages, and WhatsApp data ``whether your Gemini Apps Activity is on or off''} further eroded user control.
    \item  \textbf{Grok/xAI} requires navigating to Settings → Privacy and Safety → Data sharing and personalization → Grok to disable training, but \textit{data already processed cannot be removed from models}. The EU forced X to delete illegally processed data from May-August 2024, but this applied only to European users~\citep{rpc_grok_suspension}.
\end{itemize}

Beyond these contrived interfaces, additional concerning practices further undermine user control over their data.

\paragraph{``Temporary/vanish Chat'' mode still keeps your data for 30 days.}
Features marketed as privacy-enhancing often provide minimal actual protection. ChatGPT's ``Temporary Chat'' mode \textit{still retains conversations for up to 30 days for ``safety purposes''}~\citep{openai_temp_chat}. Google's ``incognito'' mode for Gemini similarly maintains retention for safety monitoring, while Grok's ``private'' conversations remain visible to xAI for moderation purposes~\citep{gemini_privacy_hub, xai_privacy}.

\paragraph{Deletion is an option ... only on paper and not in practice!}
Despite offering deletion options, practical limitations severely restrict user control. \textit{OpenAI's consumer users cannot actually delete their data due to the federal court order}, with the standard 30-day deletion timeline overridden by indefinite legal hold~\citep{venturebeat2025retention}. Enterprise customers with Zero Data Retention agreements remain exempt, creating a stark privacy divide~\citep{openai_enterprise}.

Anthropic's deletion excludes safety-flagged content (up to 7 years), feedback submissions (10 years), and anonymized research data. Google's deletion process takes up to 2 months with \textit{encrypted backup retention for up to 6 months}, while human-reviewed content persists for 3 years regardless of user deletion requests~\citep{anthropic_retention, gemini_privacy_hub}.

Grok faces unique challenges where \textit{data integrated into trained models cannot be removed retroactively}, and cross-platform integration with X means deletion from one service doesn't remove data from others~\citep{malik2024disable}.

\paragraph{Your paid subscription doesn't protect you, only enterprise customers get 'perfect' treatment.} 
The pricing structure of LLM services creates fundamental inequities in privacy protection. \textit{Users who cannot afford premium subscriptions must accept training on their data as the price of access}, while enterprise customers receive comprehensive protections including zero data retention and data processor agreements~\citep{openai_enterprise, anthropic_privacy_2025}. This economic discrimination is particularly acute for users in developing nations who may rely on free tiers for essential services like translation, \textit{effectively trading their linguistic data for basic functionality}~\citep{mireshghallah2024trust}. Research on datasets like WildChat reveals that many users cannot afford premium services or are geographically blocked from accessing them, making free tiers with privacy-invasive defaults their only option for accessing state-of-the-art AI capabilities.

\paragraph{Courts can override your privacy settings with a single order—and they already have.}
Legal processes completely override user privacy preferences. \textit{The New York Times lawsuit against OpenAI demonstrates how external litigation can force indefinite data retention affecting millions of users who aren't party to the case}~\citep{nyt_lawsuit_2024}. Similar court orders could affect any provider, with national security letters and law enforcement requests creating additional retention requirements that \textit{users never learn about}~\citep{google_information_requests}.

Recent incidents highlight enforcement gaps: Italy fined OpenAI €15 million for GDPR violations, \textit{over 225,000 OpenAI credentials appeared on dark web markets}, \textit{370,000+ Grok conversations were exposed through public links}, and Samsung employees leaked proprietary data through ChatGPT~\citep{italy_fine_2024, wald_incidents, gadget_grok_leak}.

\paragraph{Fun features can be data collection honeypots.}
Viral engagement features create massive data extraction opportunities. Image generation uploads, voice mode recordings, memory features that persist across conversations, and personality customization all generate comprehensive behavioral profiles~\citep{felloai_voice, surfshark_privacy}. Connected app permissions through OAuth enable data exchanges between AI chatbots and external services without clear boundaries~\citep{medium_privacy_2024}.

\textbf{Third-party browser extensions and API wrappers create additional vulnerabilities.} Popular tools like ChatGPT Writer and Merlin collect extensive user data despite privacy claims, while side-channel attacks can eavesdrop on conversations through metadata analysis~\citep{ziradaily_hackers}. Mobile apps introduce location tracking, with \textit{45\% of AI chatbot apps collecting location data}~\citep{surfshark_privacy}.

These data collection and retention policies reveal a systematic pattern of privacy erosion disguised as user choice.
In the following sections we examine how these policy failures enable broader categories of privacy violations throughout the LLM ecosystem.

\subsection{Different Types of Data in the LLM Ecosystem}
Modern LLM systems process three distinct categories of data, each presenting unique privacy challenges that current frameworks inadequately address. Understanding these data types—user interactions, system-retrieved information, and publicly available corpora—is essential for comprehending how privacy violations manifest across the LLM ecosystem.
\subsubsection{User Interaction Data \emoji{speechballoon}}\label{sec:interactions}

\textbf{User interaction data captures every digital footprint within LLM systems.} User interaction data encompasses every action users take within LLM systems: prompts typed, files uploaded, buttons clicked, voice recordings made, feedback provided, and even passive engagement metrics like session duration and feature usage patterns. Recent empirical studies reveal the deeply personal nature of this data—\cite{mireshghallah2023confaide} analyzed real-world LLM conversations and found users routinely share intimate details including mental health struggles, financial information, medical symptoms, and relationship problems, with their ConfAIde benchmark demonstrating that GPT-4 and ChatGPT inappropriately reveal private information 39\% and 57\% of the time respectively. The Washington Post's investigation into training datasets exposed how interaction data contains personal, proprietary, and offensive content collected without explicit consent~\citep{washingtonpost2023inside}, while \cite{zhang2024fair} found that users operate under false assumptions about privacy protection, particularly believing that paid subscriptions guarantee data security.

\subsubsection{System Retrieved Data \emoji{scroll}}\label{sec:rag}

\textbf{Context windows are exploding while retrieval systems access vast external data.} Modern LLM systems operate through sophisticated retrieval pipelines that access and process vast quantities of external data, with context windows experiencing explosive growth—GPT-4.1 reached 1 million tokens in January 2025 (125x growth from GPT-4's original 8,192 tokens), while Google's Gemini 1.5 Pro processes up to 2 million tokens~\citep{ibm2024context,gemini2025context}. Retrieval-Augmented Generation (RAG) systems retrieve diverse data types including textual documents (research papers, legal contracts, support tickets), structured data (database records, financial statements, spreadsheets), multimodal content (images with extracted text, video transcripts), and real-time information (current events, API responses, social media feeds)~\citep{zou2025poisonedrag}. These systems employ semantic search using vector embeddings with 256-512 token chunks, hybrid search combining sparse and dense retrieval methods, and multi-hop reasoning across multiple documents, with research showing that 4K context LLMs with RAG achieve comparable performance to 16K context fine-tuned models~\citep{f52024rag}.

\textbf{Persistent memory and tool integrations compound privacy attack surfaces.} The emergence of persistent memory architectures compounds data exposure risks through vector databases storing conversation embeddings, graph databases maintaining relationship networks, and hybrid storage systems combining structured and unstructured data~\citep{zep2024memory,packer2023memgpt}. Tool integrations further expand the attack surface by retrieving structured API responses, executing function calls with return values, accessing real-time data feeds from IoT devices and financial markets, and performing file system operations for document analysis. \textbf{The era of ``mega-contexts'' erases the line between private and shared data.} As context windows approach ``mega-contexts'' where wearable devices, smart home assistants, and personal computing environments feed continuous streams of intimate data into LLM systems, the distinction between ``shared'' and ``private'' data effectively disappears—creating an unprecedented expansion of the privacy attack surface that current frameworks fail to address~\citep{edpb2025llm,gan2024navigating}.

\subsubsection{Publicly Available Data \emoji{globewithmeridians}}\label{sec:public}

\textbf{``Public'' training data was never consented to for AI use and contains extensive personal information.} The vast corpora of publicly available data used to train LLMs present a paradox: while technically ``public,'' this data was never consented to for AI training purposes and contains extensive personal information, copyrighted material, and embedded security vulnerabilities. Training datasets are contaminated with thousands of live secrets and credentials Training datasets like Common Crawl's December 2024 archive (400TB, used by DeepSeek, OpenAI, and others) contain approximately 12,000 live API keys and passwords, including AWS root keys and Slack webhooks, with 63\% of secrets repeating across multiple pages—meaning LLMs trained on this contaminated data may inadvertently generate unsafe outputs~\citep{truffle2025secrets}. Legal frameworks are evolving rapidly, with a 2025 European paper establishing that LLMs themselves can be classified as personal data under GDPR if information extraction makes individual identification ``reasonably likely,'' while ongoing copyright litigation like \textit{New York Times v. OpenAI} and \textit{Bartz v. Anthropic} creates bifurcated frameworks where training may qualify as fair use but unlawful data acquisition still constitutes infringement~\citep{nytimes2024lawsuit,bartz2025anthropic}.
\cite{aiRobotsTxt} tracks AI-related crawlers and identified crawlers that do not respect websites' \texttt{robots.txt}.

\textbf{``Public'' inference data suffering from democratized surveillance at scale.} The democratization of sophisticated intelligence gathering through LLM-powered tools like Deep Research enables aggregation of dispersed information—deadnames, security questions, childhood addresses—at costs under \$1 per task with F1 scores above 0.94, effectively weaponizing previously obscure public data through automated synthesis and cross-platform correlation~\citep{llmosint2024,staab2024beyond}.

%%%%%%
\section{How is the Data Being Exposed?}
\label{sec:exposure}

Having established the three primary data types at risk in Section~\ref{sec:data_effect}—user interactions, system-retrieved data, and publicly available data—we now examine the mechanisms by which these data become exposed, creating privacy incidents across the LLM ecosystem. Table~\ref{tab:taxonomy} presents our comprehensive taxonomy of five distinct exposure pathways, each targeting different combinations of the data types we identified. This section systematically analyzes these exposure mechanisms, revealing how current technical and policy frameworks fail to address the multifaceted nature of modern LLM privacy threats.

\subsection{Training Data Leakage via Regurgitation}\label{sec:regurg}
This exposure pathway, outlined as the first incident type in Table~\ref{tab:taxonomy}, occurs when models act as data stores that inadvertently reveal training data to innocent users or malicious actors seeking to extract information. While this category has received disproportionate research attention, our analysis reveals important nuances between pre-training and post-training memorization risks.

\subsubsection{Verbatim Regurgitation of Pre-training Data is Overrated}

\textbf{The verbatim memorization narrative has been overstated as a privacy threat.} The verbatim memorization and exact regurgitation of pre-training data, especially data that appeared fewer than four times during training, has been extensively studied and shown \textit{not to pose significant privacy, security, or copyright risks}~\citep{carlini2021extracting}. Membership inference attacks (MIAs) and extraction attacks on pre-training data have demonstrated limited efficacy under typical modern pre-training settings. Models are usually trained on vast, openly available corpora, using large batch sizes, minimal epochs, and substantial dataset diversity, greatly diluting potential memorization effects. \cite{huang2024demystifying}'s work confirms that exact extraction typically requires non-trivial amounts of data repetition, further diminishing the real-world threat.

Another mitigating factor arises from \textit{model capacity dynamics}. Early in training, models lack linguistic proficiency, limiting memorization capabilities. Later in training, increased language proficiency paradoxically reduces memorization by efficiently encoding generalized representations rather than specific verbatim data points~\citep{huang2024demystifying}.

\subsubsection{Fine-tuning and Post-training Memorization Risks are Real}

\textbf{Post-training phases present legitimate and understudied memorization risks.} However, fine-tuning and post-training present legitimate privacy concerns, contrasting significantly with pre-training scenarios. Fine-tuning involves smaller datasets, more epochs, and stronger recency bias, conditions conducive to memorization~\citep{borkar2025privacy}. Additionally, interactions between \textit{model size, linguistic capability, and training stage} play critical roles in memorization. Mid-training, when models gain competence but remain below full capacity, creates a vulnerable phase during which memorization becomes notably efficient~\citep{borkar2025privacy}.

This phase can induce \textit{emergent misalignments}, where \textbf{non-contiguous sequences, co-occurrences, and subtle contextual interactions} result in unintended behaviors and leaks~\citep{borkar2025privacy}. Unlike literal memorization, these emergent memorization risks extend to personally identifiable information (PII) or other sensitive content that the model may unintentionally regurgitate, even if explicitly excluded from training.

For example, Ripple Effect studies highlight how fine-tuning induces memorization of nuanced sequences that lead to leakage, illustrating the overlooked complexities within fine-tuning memorization risks~\citep{borkar2025privacy}. As fine-tuning commonly includes user-provided data, such risks become particularly consequential, necessitating careful scrutiny and targeted mitigation strategies.

\subsubsection{Beyond Verbatim Leakage: Semantic, Cross-lingual, and Cross-modal Leakage}

\textbf{Modern privacy threats extend far beyond literal text regurgitation.} Emerging research further underscores leakage risks beyond literal textual regurgitation, extending to semantic, cross-lingual, and cross-modal domains.

\paragraph{Semantic Leakage}
Semantic leakage encompasses risks related to \textit{conceptual rather than literal information}. Studies utilizing non-literal copying benchmarks and semantic re-identification frameworks have demonstrated how models may leak semantic or distributional information not explicitly contained in verbatim training data~\citep{chen2024copybench}. 

\paragraph{Cross-lingual Leakage}
Cross-lingual leakage arises when information originally presented in one language leaks into outputs in another language, exploiting shared n-gram structures and conceptual overlaps across linguistic datasets. Recent works, such as those by \cite{dong2025understanding}, provide concrete evidence of how multilingual models unintentionally transfer sensitive content across languages, amplifying potential privacy risks across linguistic boundaries.

\paragraph{Cross-modal Leakage}
Cross-modal leakage represents another frontier of privacy risks, involving data memorization and leakage across different modalities. Recent \textbf{phoneme-based attacks} have demonstrated that models trained across text-audio modalities can unintentionally expose data, such as lyrics or audio cues, even without literal textual overlap~\citep{roh2025bob}. Attacks leveraging phonetic similarity, with zero literal n-gram overlap, have successfully retrieved sensitive audio data, underscoring significant vulnerabilities and necessitating further attention~\citep{roh2025bob}.

Section~\ref{sec:exp} demonstrates that nearly half (48.4\%) of all AI/ML privacy research focuses on this category, representing a significant misallocation of research effort relative to real-world threats.

\subsection{Direct Chat Leakage via Uninformed Consent or Compromised Provider}\label{sec:direct-leak}

As categorized in Table~\ref{tab:taxonomy}, this incident type involves the exposure of full user interaction transcripts through mechanisms where the model itself is not directly involved—rather, the vulnerability lies in the surrounding infrastructure and policies. These leakages can expose data to innocent bystanders, legal proceedings, or malicious third parties through provider-level failures.

Beyond the memorization risks involved when user prompts are used to train models, we want to further highlight real risks of exposure that occur through LLM providers.

\subsubsection{Heightened Risks of Security Breaches in Centrally Hosted Models}

\textbf{Centralized data collection creates unprecedented attack surfaces with massive sensitive data stores.} LLM prompt data has become a highly sensitive type of information, given its widespread and frequent use across a wide range of personal and sensitive domains~\citep{mireshghallah2024trust}. This sensitivity is compounded by the fact that massive amounts of such data have been collected and stored by various types of model providers, from flagship LLM service providers and major companies hosting (proprietary) in-house models (such as OpenAI, Anthropic, DeepSeek, Google, and Meta), to niche companies (with many being startups)~\citep{wang2025large} and research labs~\citep{hou2025insecurity} running services powered by self-hosted open-source models, which may be less stable and invest less in security protection.

The risk deserves increased attention due to multiple real-world data breach incidents. In July 2025, a security flaw in Meta AI's chatbot has been reported which allowed users to access and view private prompts and AI-generated responses from other users~\citep{metaai2025breach}. In January 2025, Wiz Research discovered a publicly accessible database belonging to DeepSeek that allowed full control over database operations, including access to internal data. The exposure included over a million lines of log streams containing chat history, secret keys, backend details, and other highly sensitive information~\citep{theori2025deepseek}.
The OmniGPT breach in February 2025 compromised 34+ million user messages and 30,000+ accounts~\citep{forcepoint2025omnigpt}. These breaches create cascading harms: financial losses from API hijacking reaching \$100,000 daily, regulatory penalties under GDPR and CCPA, reputational damage to organizations whose employees leaked proprietary information, and national security concerns leading the U.S. Navy and House of Representatives to ban DeepSeek from government networks~\citep{nsfocus2025invisible}.

Given the trend of users increasingly relying on centrally hosted models in exchange for convenient access to the most performant systems, there is a growing concern about potential future data leakage incidents that could pose significant risks to individuals and businesses and even trigger a broader societal trust crisis.

\subsubsection{Hidden Agreements and Power Asymmetries in Privacy Policies}

\textbf{Privacy policies systematically favor data collection through deceptive design and power imbalances.} In addition to security breaches, model providers often have data exposure specified in the privacy policies that can be unknowingly agreed to by consumers. For example, OpenAI specifies that they use de-identified Personal Data to ``analyze the way our Services are being used, to improve and add features to them, and to conduct research,'' and ``use Content you provide us to improve our Services, for example to train the models that power ChatGPT''~\citep{openai_privacy_policy}. Gemini clearly indicates that human reviewers will ``read, annotate, and process your Gemini Apps conversations''~\citep{gemini_privacy_hub}. They shift the burden to users and expect privacy concerns to be addressed primarily through users' self-censorship. However, ~\cite{zhang2024fair} found that this expectation is unrealistic, as it significantly compromises the convenience and utility of the service, ultimately nudging users to accept data collection and sacrifice their privacy.

Due to unclear design and potential dark patterns, users' conversations with LLMs may be exposed more widely than they expect---as shown in a recent news article that Google is indexing conversations with ChatGPT that users have shared with others, turning private exchanges intended for small groups into search results visible to millions~\citep{fastcompany2025indexing}.
Grok users sharing conversations via a button inadvertently made 370,000+ conversations publicly searchable online~\citep{gadget_grok_leak}.

Another layer of intransparency may result in further unexpected data exposure from ``LLM wrapper'' apps that call APIs from model providers. The app developers can choose to share API inputs and outputs with OpenAI, with programs offering ``daily complimentary tokens'' for traffic shared~\citep{openai2024feedback}. OpenRouter.ai has offered free models that log all prompts and completions~\citep{openrouter2024logging}. However, the actual data subjects typically have no way to know about, control, or benefit from the data sharing or monetary incentives.

\subsubsection{Legal Risks}

\textbf{LLM conversations lack professional privilege protections, creating legal vulnerabilities.} People use LLMs for tasks they do not want to share with humans, including mental health support, legal advice, and health inquiries. In real life, professions that handle sensitive information are bound by legal confidentiality. However, such protections have not been established for LLMs, putting consumers' privacy at risk in the face of legal subpoenas or use as evidence in lawsuits. For example, OpenAI has been contesting a court order in its lawsuit with The New York Times that would require it to retain the chat histories of hundreds of millions of ChatGPT users worldwide~\citep{openai2024nyt, venturebeat2025retention}.

Section~\ref{sec:exp} demonstrates that 43.6\% of all AI/ML privacy research focuses on this category, while the research effort skews towards private or decentralized learning/inference techologies, which still falls short for tackling real-world threats related to the uninformed consent and legal risks in the increasingly prevalent centralized data collection regime.

\subsection{Indirect Chat and Context Leakage via Input-Output Flow}\label{sec:indirect-leak}

{This third category of incidents in Table~\ref{tab:taxonomy} emerges when models operate as autonomous agents, processing user interactions and retrieved documents through tools and APIs, creating new vectors for data exposure to malicious actors or innocent bystanders.} The expanded capabilities of modern LLM systems introduce privacy risks that extend far beyond traditional chat paradigms.

Beyond traditional chatbot interactions, modern LLM systems increasingly operate on external data through retrieval mechanisms and execute real-world actions via tool integrations. This expanded capability surface introduces new privacy leakage vectors that extend far beyond the direct chat paradigm.

\subsubsection{Risks of RAG Systems}

\paragraph{Adversarial Attacks}
RAG (Retrieval-Augmented Generation)~\citep{lewis2020retrieval} systems create new targets for data extraction attacks, demonstrated as feasible via prompt injection~\citep{zou2025poisonedrag} as well as data poisoning during training that inject backdoors to LLMs~\citep{peng2024data}. The retrieved data can be further leaked through integrated tools (e.g., sending emails).

\paragraph{Side effects of personalization.}
\textbf{Memory features create intimate surveillance that users cannot fully control or comprehend.} Many LLMs provide memory capabilities to personalize response generation, such as ChatGPT (OpenAI), Gemini (Google), Microsoft Copilot, and Grok (xAI)~\citep{google2025memory, microsoft2025memory, techcrunch2025grok}. This feature presents practical threats because: (1) users often cannot remember all information they've entered, leading to perceptions that ``ChatGPT knows more about me than I do''---many find this unsettling~\citep{reddit2024memory}; (2) The generation process may not fully understand context to determine whether personalization is appropriate, and various output channels (careless copy-paste, web search, voice mode) increase unintended data leakage risk. For example, when using voice mode, the model might speak a response containing private details in public~\citep{xiaohongshu2024voice}.

\subsubsection{Agent-Specific Risks}

\textbf{Autonomous agents amplify privacy risks through elevated permissions and minimal oversight.} LLM agents leverage capabilities such as planning, memory, and tool use. These agents are rapidly emerging, including GUI-based agents (Computer Use Agent, ChatGPT Agent, Manus.ai) and terminal-based agents (Cursor, Claude Code). Their high autonomy and open-ended functionality make it substantially more difficult to predict and control privacy leakage potential. These risks arise both in the presence and absence of malicious attackers from three key capabilities: access to private data, exposure to untrusted content, and ability to communicate externally~\citep{willison2025trifecta}. When combined, these create powerful attack vectors. One example is the recent Supabase MCP leak incident involving prompt injection where a malicious user tricked an LLM agent (e.g., Cursor) connected via Supabase MCP with \texttt{service\_role} privileges into reading private data and writing that information back into the ticket, effectively exposing the entire SQL database~\citep{generalanalysis2025supabase}.

\paragraph{LLMs lack contextual privacy capabilities.}
\textbf{Current LLMs cannot reliably make context-appropriate privacy decisions.} Prior research~\citep{mireshghallah2023confaide} has shown that LLMs have limited capabilities for making appropriate privacy-related decisions given context. Contextual Integrity (CI)~\citep{nissenbaum2009privacy} posits that data flow appropriateness is context-dependent and governed by norms specified through five key parameters: data sender, subject, recipient, type, and transmission principles. The extent to which LLMs possess these capabilities remains uncertain, making it premature to reliably integrate LLM agents into open-ended environments with full access to our social lives.

\paragraph{Overburdening users with privacy control}
Current agentic systems rely on users as the last resort, expecting them to carefully monitor the agent's actions to prevent harms (e.g., OpenAI Operator~\citep{openai2025introducing}) and to actively delete external data exposed to the agent through tool use (e.g., Connectors, see \autoref{fig:connectors}).
However, rudimentary privacy control designs often fall short in both overcoming human cognitive limitations in identifying privacy violations and doing so without causing unnecessary disruption.
A paradox seems to have emerged: users need to feel that they retain final authority over agents' actions to build trust, yet human oversight has been found to be largely ineffective at identifying and preventing privacy harms~\citep{zhang2024oversight, chen2025obvious, tang2025dark}.
% Additionally, situations where data sharing appropriateness is inherently ambiguous should require human input~\cite{li2025privaci}. 

As we demonstrate in Section~\ref{sec:exp}, research on these agent-based privacy risks remains critically understudied, representing only 2.0\% of published work despite their rapidly growing real-world deployment.

\subsection{Privacy Under the Microscope: Indirect Attribute Inference}\label{sec:micrs}

{While previous exposure mechanisms focused on direct data leakage, this incident and the next incident types in Table~\ref{tab:taxonomy} represent a fundamentally different privacy threat: the use of LLMs as inference and search engines to extract or aggregate sensitive attributes about bystanders from available data.} These capabilities democratize sophisticated surveillance and inference attacks, enabling malicious users to violate privacy at unprecedented scale.

\textbf{LLMs enable sophisticated inference attacks that extract sensitive attributes from seemingly innocent data.} LLMs can be exploited as privacy inference engines, deriving location, occupation, or ethnicity from ordinary conversation without direct identifiers~\citep{staab2024beyond}, or inferring geolocations from seemingly ordinary images~\citep{mendes2024granular}. In a viral social-media trend reported in mid-April 2025, users uploaded photos as innocuous as dimly lit bars or random street corners to ChatGPT (using new o3 and o4-mini models), and the model quickly and often correctly identified locations---raising real-world doxxing and privacy concerns~\citep{techcrunch2025geolocation}. Participants in a Hacker News thread described the capability as ``surreal, dystopian and entertaining,'' with one remarking: ``Accessible to anyone, superhuman levels of deductive reasoning to pick out your location from super minor details in an innocent photo? That could certainly be dystopian''~\citep{hackernews2025geolocation}.

As we demonstrate in Section~\ref{sec:exp}, research on indirect attribute inference privacy risks remains understudied, accounting for only 5.8\% of published work.
Despite this already low number, a significant portion focuses only on pre-LLM versions of the problem, such as inferring sensitive attributes from text embeddings, which differ in scope of impact and require distinct mitigation methods.

\subsection{Privacy Through the Telescope: Direct Attribute Aggregation}\label{sec:teles}

\textbf{Agentic search capabilities democratize surveillance by lowering barriers to comprehensive data aggregation.} The public internet faces unprecedented privacy threats as agentic capabilities---such as Deep Research in ChatGPT~\citep{openai2025research}---drastically lower the barrier to aggregating, synthesizing, and analyzing large volumes of online information. This empowers legitimate use cases but also gives non-technical users unprecedented power to dig up sensitive details, enabling cyberstalking, doxxing, and impersonation. Anecdotal evidence reveals sensitive information exposure such as pets' names (often used for security questions, see Figure~\ref{fig:pet_cat}) or deadnames of transgender persons, creating risks of account hacking, targeted scams, emotional distress, and discrimination~\citep{liu2025evaluating,kim2025llms}.  This threat extends beyond privacy into security, as seemingly innocuous information can be exploited to steal accounts through secondary questions.

The risks are heightened when LLM-powered search integrates with closed systems like social media. The AI-powered search feature on Weibo (256 million daily active users~\citep{weibo2025stats}) works as a RAG system, retrieving users' posts and summarizing them using the DeepSeek-R1 model. In April 2025, Chinese netizens discovered that searching user IDs could lead to unwanted exposure of personal details, with suspicions that even private posts might be included, sparking heated discussion and widespread panic~\citep{weibo2025incident}.

As we demonstrate in Section~\ref{sec:exp}, research on direct attribute aggregation privacy risks remains critically understudied, accounting for only 0.2\% of published work.

\section{A Decade of AI/ML Privacy Research: Trends from Leading ML, NLP, and S\&P Conferences}\label{sec:exp}

Having established a taxonomy of five distinct privacy incident types in LLM systems, we now examine how the research community has addressed these threats. We analyze 1,322 AI/ML privacy papers published at top conferences from 2016--2025, mapping them to our incident categories to identify gaps between research focus and real-world privacy risks.

\subsection{Corpus}

We collect a comprehensive corpus of papers from top ML, NLP, and S\&P conferences published between 2016 and 2025.
We opted to use a ten-year window to allow for a longitudinal trend analysis.
Also, 2016 is a critical time point when the original paper on DP-SGD~\citep{abadi2016deep} was published.
Relatedly, the original paper on Federated Learning was published in 2017~\citep{mcmahan2017communication}.
We believe this selection ensures a decent coverage of technical privacy research on modern machine learning (e.g., deep learning, large language models).

We select three top security and privacy conferences, which are USENIX Security, IEEE S\&P, and ACM CCS.
We select three top AI/ML conferences, which are ICML, ICLR, and NeurIPS.
We also select two top NLP conference, ACL and EMNLP, as LLM is our primary focus of analysis.
The three security conferences, ICML, and NeurIPS were scraped from the official proceedings websites.
The ACL anthology was downloaded directly from \url{https://aclanthology.org/}.
For ICLR, we use an existing dataset \url{https://github.com/berenslab/iclr-dataset}.

\subsection{Annotation Pipeline}

\subsubsection{AI/ML privacy paper filter} We filter papers on AI/ML privacy.
The paper must involve modern AI/ML technologies such as deep learning and large language models. We exclude traditional ML methods such as logistic regression.
The paper must study privacy issues related to AI models or systems rather than using AI to solve general security problems.
We develop a prompt to annotate the data with the GPT-4.1 model.
One author experienced in reviewing and publishing S\&P papers labeled 50 papers sampled from all the papers from the S\&P conferences to evaluate the annotation pipeline, achieving an accuracy of 100\%.
Another author experienced in ML/NLP labeled 50 papers sampled from the ML conferences, achieving an accuracy of 96\% (Cohen's kappa=0.90, Gwet's AC1=0.93); 
and also labeled 50 papers sampled from the NLP conferences, achieving an accuracy of 100\%.
Our classifier identified 1,322 AI/ML privacy papers from the corpus.

\subsubsection{Target Incident Classification}

Finally, we analyze the AI/ML privacy papers in our dataset to align them with the five types of personal data incidents in LLM systems.
We translate each type of incident into research topics as shown in \autoref{tab:personal-data-incidents}.
We then develop a prompt to annotate the data with the GPT-4.1 model.
The prompt was applied to small samples of the data and improved iteratively.
The final evaluation was conducted on a sample of 50 AI/ML privacy papers and achieved an accuracy of 96\% (Cohen's kappa=0.93, Gwet's AC1=0.94).

\begin{table}[htbp]
\centering
\caption{Mapping of Personal Data Incidents in Large Language Model Systems to Research Topics}
\label{tab:personal-data-incidents}
\resizebox{\textwidth}{!}{%
\begin{tabular}{p{5cm}p{8cm}}
\toprule
\textbf{Incident Type} & \textbf{Research Topics} \\
\midrule
Training Data Leakage via Regurgitation & 
Membership Inference Attack, Attribute Inference Attack, Data Extraction Attack, Model theft and extraction Attacks~\citep{tramer2016stealing}, Differentially Private Model Training, Machine Unlearning
\\
\midrule
Direct Chat Leakage via Uninformed Consent or Compromised Provider & Audit collection of prompts; Side channels that allow prompt leakage; Private inference and training that avoids centralized collection of raw data, including on-device inference/training, Homomorphic Encryption (HE), secure multi-party computation (MPC), federated learning (FL), Trusted Execution Environments (TEEs) \\ 
\midrule
Indirect Chat and Context Leakage via Input-Output Flow & Contextual Integrity, Prompt Injection, Leakage from In-Context Learning, RAG, and Agentic AI, etc. \\
\midrule
Indirect Attribute Inference & Image Geolocalization, User Profiling, And countermeasures to avoid inference of identity or attributes, etc. \\
\midrule
Direct Attribute Aggregation & Extracting Personal Information from Public Data, CyberAttacks etc.\\
\bottomrule
\end{tabular}
}%
\end{table}

\subsection{Results}
\label{sec:papertrends}

\begin{figure}
    \centering
    \includegraphics[width=1\linewidth]{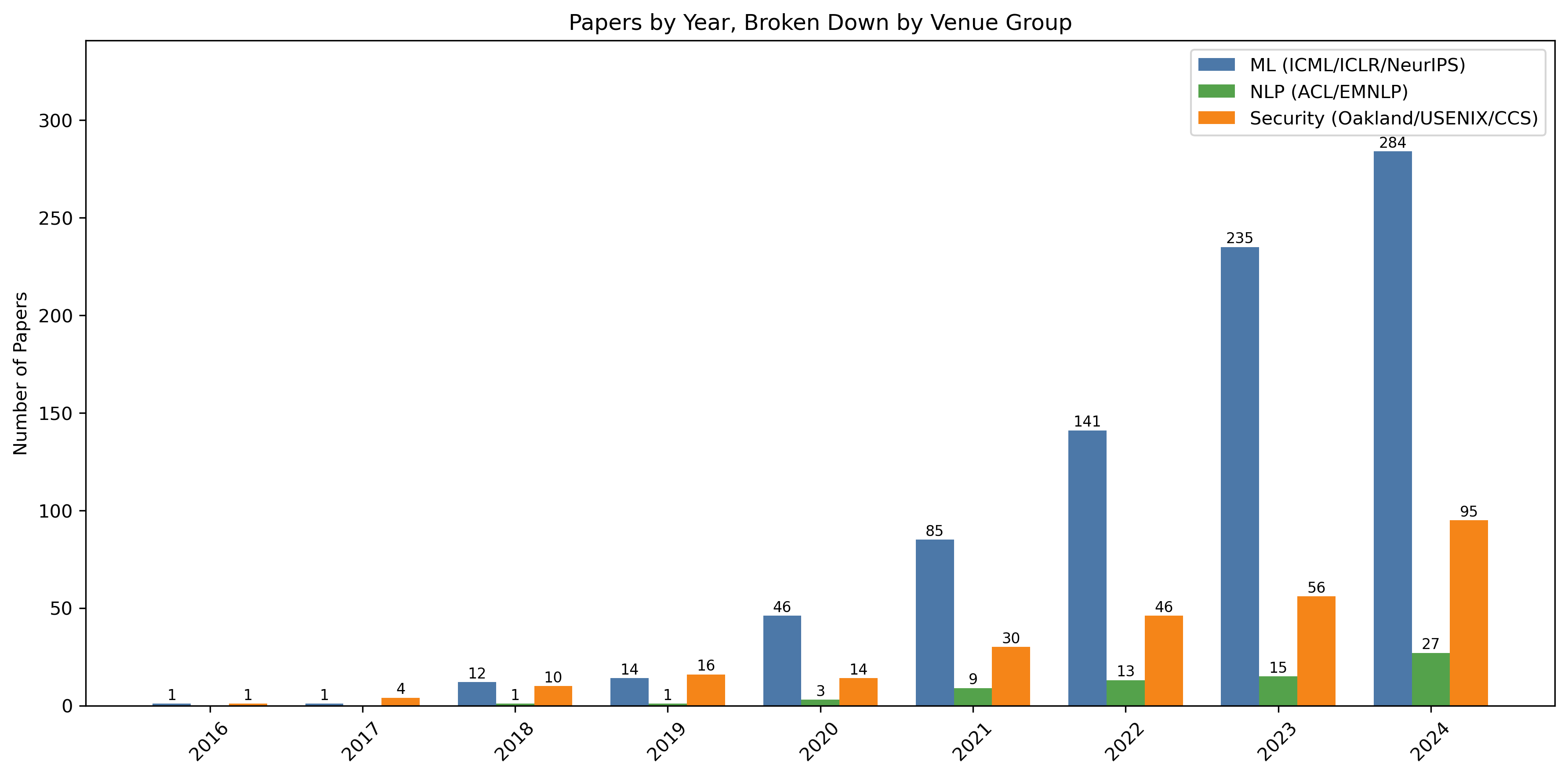}
    \caption{AI/ML Privacy Papers by Years, Broken Down by Venue Group}
    \label{fig:papers_by_year_by_venue_group}
\end{figure}

\begin{figure}[h!]
    \centering
    \begin{minipage}{0.48\linewidth}
        \centering
        \includegraphics[width=\linewidth]{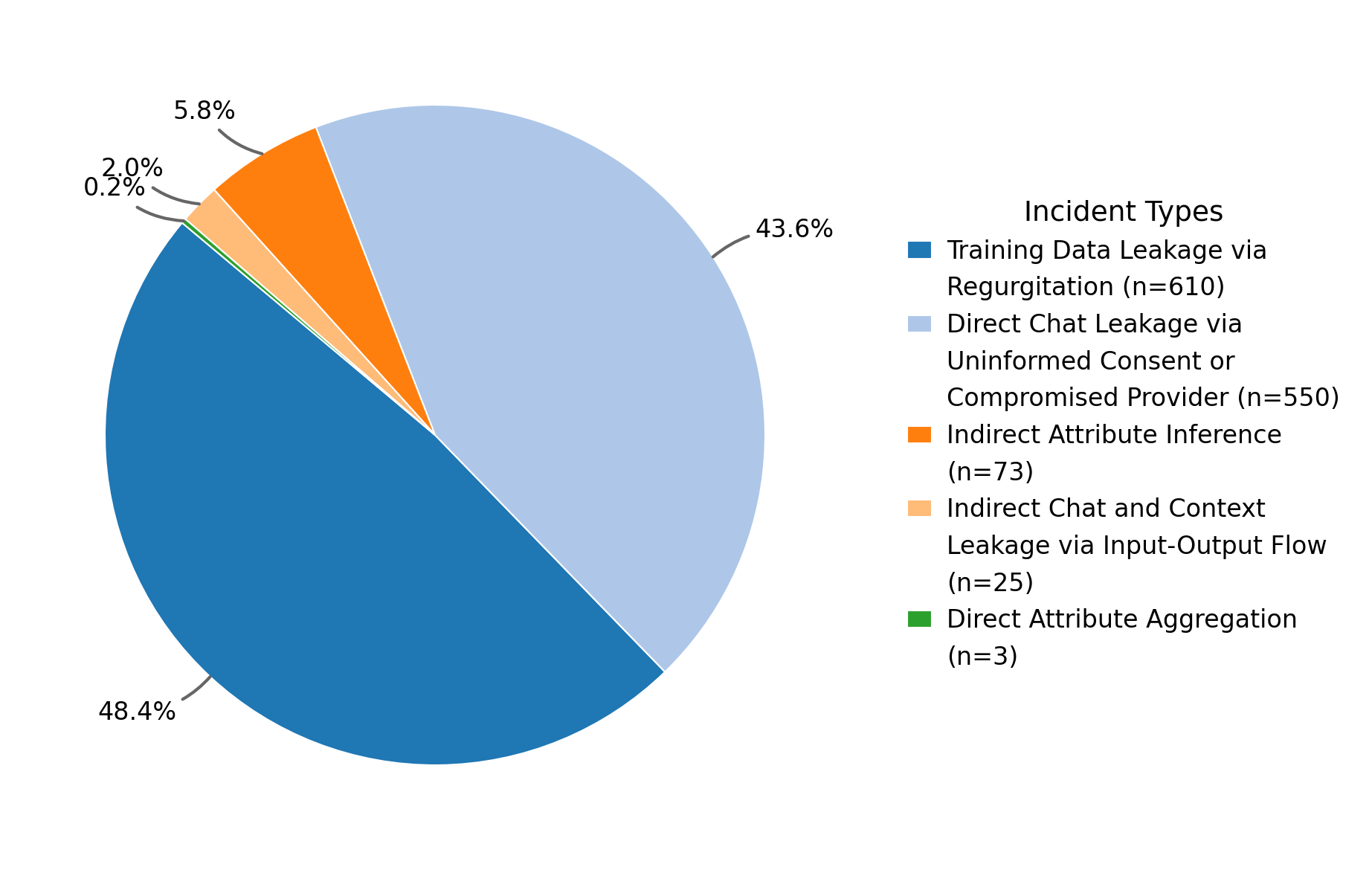}
        \caption{Incident type distribution}
        \label{fig:incident_type_distribution}
    \end{minipage}%
    \hfill
    \begin{minipage}{0.48\linewidth}
        \centering
        \includegraphics[width=\linewidth]{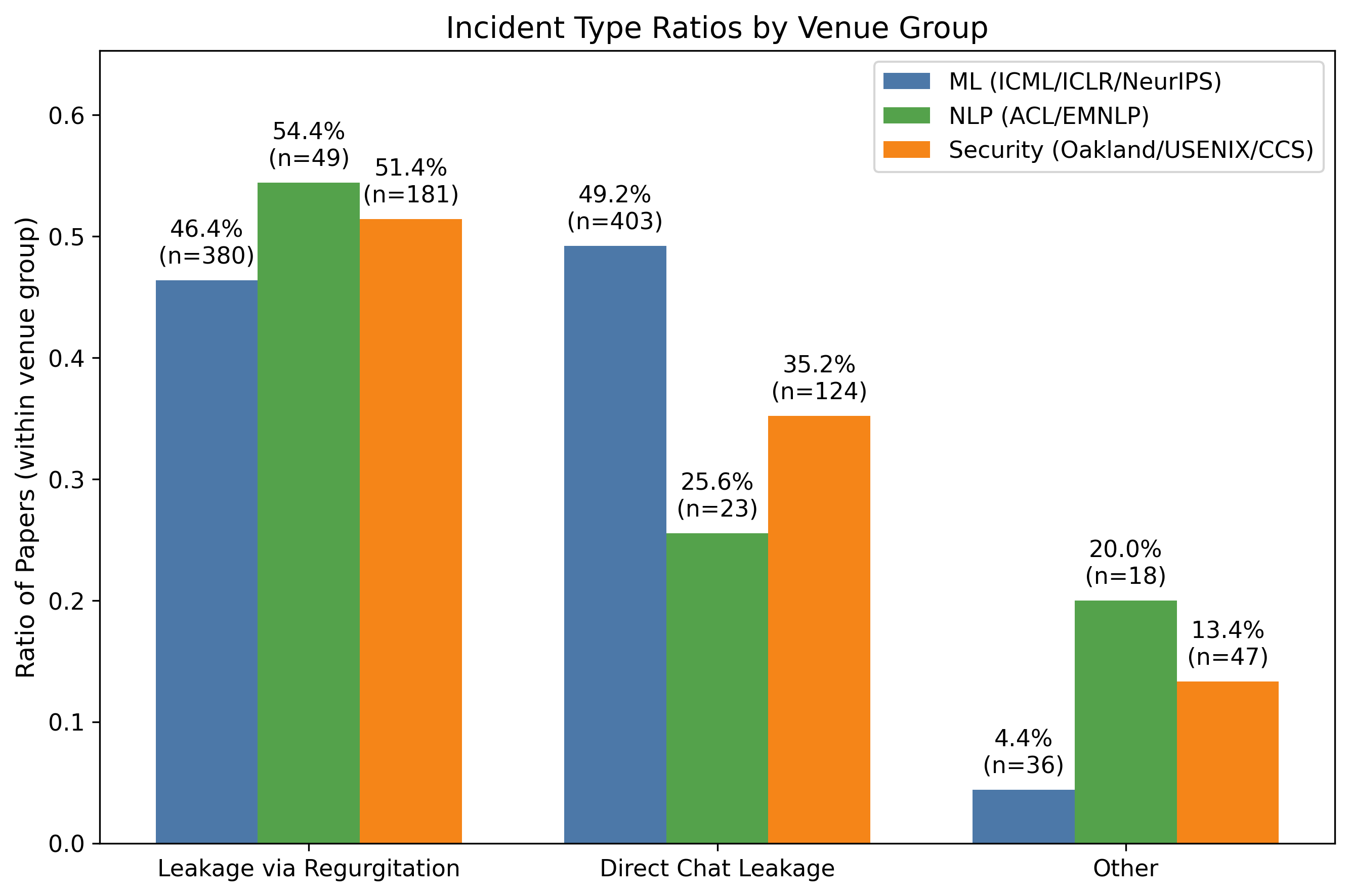}
        \caption{Incident Type by Venue Group (ML, NLP, Security conferences)}
        \label{fig:incident_type_by_venue_group}
    \end{minipage}
\end{figure}

We observe a strong upward trend in research on AI/ML privacy since 2016 across ML, NLP, and Security venues (see \autoref{fig:papers_by_year_by_venue_group}).
However, when translating the potential real-world impact of this research through the lens of the incident types it can identify or mitigate (\autoref{fig:incident_type_distribution}), two categories dominate the results: Training Data Leakage via Regurgitation (48.4\%) and Direct Chat Leakage via Uninformed Consent or Compromised Provider (43.6\%), together accounting for 92\% of all papers. In contrast, the other incident types, namely Indirect Attribute Inference (5.8\%), Indirect Chat and Context Leakage via Input-Output Flow (2.0\%), and Direct Attribute Aggregation (0.2\%), remain significantly understudied.

We argue that the prevalence of the two dominant categories stems from the well-developed and still growing communities surrounding certain technologies, including Differential Privacy (DP), Federated Learning (FL), Homomorphic Encryption (HE), Secure Multi-Party Computation (MPC), Trusted Execution Environments (TEE), and On-Device ML.
Except for DP, these technologies mainly limit the data sharing with a centralized server for training and inference purposes.
Conceptually, these approaches offer potential solutions to fundamentally address direct chat leakage issues. For example, homomorphic encryption (HE) could enable inference with encrypted text, ensuring that no user chat logs are exposed in the event of a security breach. Similarly, federated learning (FL) could allow model training without requiring user data to be shared with a central server, eliminating the need for dark patterns or hidden agreements in policies to coerce users into contributing data for model improvement. Running models entirely on-device further reduces concerns about sharing data with a central server.
However, in practice, these methods can introduce costs to performance and usability, sometimes prohibitively so. They may also create safety and abuse concerns---the lack of visibility into real-world AI usage can increase the likelihood of the other incidents we identified.
\textbf{As centralized data collection in LLM services has become, and will likely remain, the mainstream, there is a need for technologies that address such incidents without assuming extreme decentralization or strictly local training and inference.
}

In the third largest category, Indirect Attribute Inference (5.8\%), we observed two distinct generations of work.
Before 2024, research in this category primarily focused on learning privacy-preserving neural representations—for example, preventing the inference of sensitive attributes from text embeddings \cite{pan2020privacy} (Oakland 2020).
Since 2024, however, the rise of large language models (LLMs) has significantly expanded the attack surface.
As demonstrated by \citet{staab2024beyond} (ICLR 2024), pretrained LLMs possess strong capabilities to infer personal attributes directly from text.
This latter line of work highlights a severe concern: \textbf{such attacks have become democratized, enabling individuals with little technical expertise to perform them, and posing broader risks since text is far more prevalent in daily life than specialized neural representations.}

Finally, we would like to emphasize the differences across venues—the skewed distributions are particularly pronounced in ML conferences, where only 4.4\% of papers address the last three incident types. \textbf{By contrast, this ratio rises to 20\% in NLP conferences and 13.4\% in Security conferences. We believe it is crucial to give greater attention to these areas, which have received far less spotlight than mainstream ML research.}

\section{Technical Solutions and Beyond: A Roadmap Forward}

The privacy challenges we have identified---from power asymmetries to emergent memorization behaviors---demand a comprehensive response that spans technical, sociotechnical, and policy interventions. While no single approach can fully address the multifaceted nature of these risks, a layered defense strategy combining immediate practical solutions with longer-term research directions offers a viable path forward. In this section, we first examine technical interventions that users and developers can deploy today, including local data minimization, hybrid architectures, and privacy-aligned post-training. We then explore sociotechnical approaches that reshape the relationship between users and LLM providers through transparency, user empowerment, and community governance. Finally, we consider the policy landscape necessary to establish meaningful privacy protections in an era of increasingly capable AI systems.

\subsection{Technical Interventions}

\textbf{Local data minimization.} Systems like Rescriber demonstrate that smaller LLMs running on-device can effectively sanitize personal information before transmission to cloud services~\citep{zhou2025rescriber}. This browser extension, powered by Llama3-8B locally, achieves performance comparable to GPT-4o while maintaining complete user control over privacy-utility tradeoffs—critically important given users' documented struggles with understanding privacy implications of their LLM interactions~\citep{zhang2024oversight}. The approach addresses the power asymmetries identified in Section~\ref{sec:direct-leak} by eliminating dependence on centralized providers for privacy protection. Furthermore, \citet{dou2023reducing} (ACL 2024) demonstrate that lightweight models can be used effectively as disclosure management tools, helping individuals rephrase or moderate their own messages before posting them online, thus reducing privacy risks in online self-disclosure~\citep{10.1145/3711029}.

\textbf{On-device inference.} Modern smartphones support 7B parameter models at acceptable performance levels, while WebLLM enables high-performance browser-native inference using WebGPU acceleration~\citep{ruan2024webllm}. Browser extensions like PRISMe analyze privacy policies in real-time using local models~\citep{freiberger2025prisme}, processing data entirely on-device without requiring users to trust centralized providers with sensitive information. These tools represent a fundamental shift in the privacy-utility calculus, offering users meaningful alternatives to cloud-dependent services.

\textbf{Hybrid remote-local architectures.} Building on the Socratic Models framework~, recent work demonstrates how privacy-preserving chain-of-thought reasoning can split tasks between generic remote processing and encrypted local database searches~\citep{bae2025privacy}. The Split-N-Denoise architecture provides local differential privacy guarantees while maintaining superior privacy-utility tradeoffs through calibrated noise injection and client-side denoising~\citep{mai2023split}. Such approaches enable users to benefit from powerful cloud models while retaining cryptographic privacy guarantees for their sensitive data.

\textbf{Privacy alignment.} Constitutional AI has been extended explicitly for privacy protection, with Anthropic's framework incorporating principles derived from human rights declarations~\citep{bai2022constitutional}. The PROPS (Progressive Private Self-alignment) mechanism demonstrates that protecting only human preferences rather than entire training examples can achieve competitive performance with reduced perturbation requirements~\citep{teku2025props}. Google's research on user-level differential privacy for fine-tuning shows that production-viable privacy protection is achievable at scale, though with non-trivial computational overhead~\citep{privacy_finetuning2025}.

\textbf{Restricting model misuse.} Complementing privacy alignment efforts, \citet{deng2024sophon} (Oakland 2024) propose Sophon, a non-fine-tunable learning method designed to restrict task transferability. By structurally limiting the ability of pretrained models to adapt to unintended downstream tasks, Sophon reduces the risk of repurposing models for malicious use. Similarly, \citet{mendes2024granular} (EMNLP 2024) introduce techniques for granular privacy control in geolocation sharing, leveraging vision-language models to enforce fine-grained user-defined rules.

\textbf{Memorization vulnerabilities.} While verbatim memorization of pre-training data poses limited privacy risks, fine-tuning typically increases memorization rates from 0-5\% baseline to 60-75\%~\citep{ramakrishnan2025assessing}. More concerning are subliminal learning patterns that transmit behavioral traits through semantically unrelated statistical patterns~\citep{cloud2025subliminal}, creating hidden channels for information leakage. When combined with out-of-context reasoning capabilities~\citep{berglund2023taken} and phoneme-based cross-modal memorization attacks~\citep{roh2025bob}, these vulnerabilities enable sophisticated privacy violations through seemingly benign queries.

\textbf{Auditing adversarial capabilities.} Parallel to defensive measures, systematic auditing of LLM adversarial capabilities has become critical. \citet{liu2025evaluating} (USENIX Security 2025) benchmark the ability of LLMs to extract personal information and evaluate the efficacy of different countermeasures, shedding light on both the magnitude of the risk and the limitations of existing defenses. \citet{kim2025llms} (USENIX Security 2025) examine the agentic dimension, showing that once LLMs are equipped with web-based tools, the threat landscape expands: agents not only become more potent in executing cyberattacks but also lower the barrier to entry. \citet{zhan2025malicious} (USENIX Security 2025) demonstrate how malicious conversational AI systems can deliberately manipulate users into revealing sensitive personal information, underscoring the real-world risks of adversarial LLM deployments.

\textbf{Emergent misalignment.} Fine-tuning on narrow tasks can produce broad behavioral changes across unrelated domains~\citep{betley2025emergent}, suggesting that memorization enables conditional behaviors triggerable across diverse contexts. These vulnerability patterns persist even after heavy data filtering, creating model-specific signatures that adversaries can exploit. This fundamentally challenges our ability to predict or control privacy risks through traditional analysis of training data alone.

\textbf{Multi-layered defense.} Research demonstrates that four-layer defense—semantic deduplication, differential privacy generation, entropy-based filtering, and pattern-based content filtering—can achieve near-complete data leakage elimination while maintaining 94.7\% of original utility. Multi-agent privacy frameworks achieve 18-19\% reduction in private information leakage through specialized reasoning decomposition~\citep{li20251}, while user-led systems show no accuracy loss with improved user satisfaction~\citep{zhou2025rescriber}.

\textbf{Deployment recommendations.} We recommend: (1) implementing user-led data minimization by default with clear privacy-utility visualization, (2) providing local inference options for privacy-sensitive use cases, (3) adopting hybrid architectures that preserve cryptographic guarantees while leveraging cloud capabilities, and (4) incorporating privacy-specific alignment during post-training. Longer-term research must address the fundamental challenge of emergent memorization behaviors that create exploitable vulnerability patterns beyond the reach of current protective mechanisms.

\subsection{Sociotechnical Approaches}

Privacy is, by nature, a sociotechnical problem. New challenges often arise from technologies that enhance our ability to collect, store, analyze, and distribute information. As society adapts to these technologies, harms are inflicted on individuals, and humans must develop new practices and understandings of the world in order to remain in control of their privacy.

We have analyzed research from technical domains, and we argue that it is a myth to assume that privacy issues in AI models can or should be addressed solely within AI research.
While technical solutions are necessary, they are not sufficient: addressing AI privacy problems also requires sociotechnical approaches to ensure that solutions align with social norms and create a positive societal impact.

We will discuss the intersection with human-centered research (e.g., work published in security/usable security venues and HCI venues), which tends to focus on human problems.
The challenges and opportunities lie in translating these problems for the technical community so that model- and system-level approaches provide fundamental capabilities, while also informing the HCI and broader design and social science communities to leverage these capabilities in designing human-centered mitigations and studying their impact on individuals, communities, and society as a whole.

\paragraph{Input Privacy Control: Repairing Awareness and Agency}

Prior work has shown that users often hold flawed mental models about how their data is used in both response generation (inference) and model improvement (training)~\citep{zhang2024fair}. This aligns with our analysis of unexpected data sources and the added complications of features such as ChatGPT's memory, where the user thinks the system ``know more about me than I do,'' as well as the indirect inference and direct aggregation threats to any online data.

People (both direct LLM users and bystanders) need better support for awareness at multiple levels: (1) what they have shared, directly or indirectly, that could be supplied to LLMs; (2) what sensitive attributes are included; (3) how this information will be used; (4) what information is memorized---whether stored, used as ongoing context, or internalized into the model; and (5) what risks or harms may result.

Recent tools illustrate promising directions. Rescriber~\citep{zhou2025rescriber} enables user-led data minimization by detecting and highlighting potentially sensitive content in user inputs, giving people greater control over sanitization. Participants reported that simply being able to see which parts of their messages were flagged as sensitive was already highly valuable. MemoAnalyzer~\citep{zhang2024ghost} offers a user-centered interface that visualizes and allows management of ChatGPT memories, thereby helping users proactively identify and resolve privacy leakages.

\paragraph{Output Privacy Control: Human Oversight in Agentic AI}

As autonomous AI agents rapidly advance and gain traction, addressing Indirect Chat and Context Leakage via Input-Output Flow incidents requires effective output privacy controls. Research has shown that human overreliance on AI can diminish the effectiveness of human oversight in ensuring privacy protection~\citep{chen2025obvious, zhang2024oversight}. This calls for further work to examine differences in the saliency of information for humans versus models, to model human errors and cognitive biases, and to design mechanisms that help people recognize their mistakes and make more rational decisions.

\paragraph{Contextual Privacy: Laws, Social Norms, and Individual Preferences}

While Contextual Integrity provides a valuable framework, it remains difficult to operationalize in practice. A growing body of work has framed privacy risks of LLMs through this lens. \citet{mireshghallah2023confaide} (ICLR 2024) introduce ConfAIde, a benchmark designed to test instruction-tuned LLMs' ability to reason about privacy in context. Their results highlight a critical gap: while models may detect direct disclosures of sensitive attributes, they frequently fail to respect contextual norms, revealing a deeper weakness in LLM privacy reasoning. Building on this foundation, \citet{li2025privaci} (ACL 2025) present PrivaCI-Bench, which evaluates privacy compliance more comprehensively. Unlike prior benchmarks focused narrowly on PII detection, PrivaCI-Bench incorporates social contexts derived from privacy laws, real court cases, and policy documents. This extension enables systematic evaluation of whether LLMs uphold legally grounded privacy norms.
\citet{fan2024goldcoin} (EMNLP 2024) propose GoldCoin, a framework that grounds LLMs in legal reasoning using Contextual Integrity. By generating synthetic judicial scenarios informed by privacy laws such as HIPAA, GoldCoin trains LLMs to detect violations across both synthetic and real-world cases. Their experiments show that models trained with GoldCoin achieve 8–23\% higher accuracy than baselines on judicial judgments and privacy-risk detection tasks, demonstrating the value of grounding contextual privacy reasoning in legal norms.

At the system level, \citet{bagdasarian2024airgapagent} (CCS 2024) address a concrete attack vector known as context hijacking, where malicious third parties attempt to manipulate a conversational agent into leaking private data. They propose AirGapAgent, a defense mechanism that enforces contextual restrictions by ensuring only task-relevant information is accessible to the agent. Whereas baseline agents' protections collapse under adversarial prompting, AirGapAgent maintains consistently high levels of privacy protection, illustrating how Contextual Integrity can guide effective system-level defenses.

However, privacy management involves multiple, sometimes conflicting, facets that extend beyond norms alone---laws, social expectations, and individual preferences all play important roles. This raises open questions: how can their differences be reconciled, and under what conditions should one facet take precedence?

One critical position we want to make is that \textbf{privacy should be studied more on the ground}. In other words, while theories provide frameworks and laws and policies establish guidelines, they remain insufficient to capture real-world nuances or fully align with actual human needs. When conflicts arise, real-world human needs should be prioritized, which requires improved elicitation methods~\citep{guoboundary2025, guo2025privi}. Legal requirements are relatively explicit, but unspoken social norms are harder to capture, and human preferences are heterogeneous, varying across individuals, contexts, and even within the same person depending on timing and stimuli.
Current resources remain limited, with only a few efforts such as ConfAIde~\citep{mireshghallah2023confaide} and PrivacyLens~\citep{shao2024privacylens}, both remain at the laws and social norms level.
What is needed are scalable, authentic, consequence-aware, and socially meaningful methods to elicit preferences and norms in context.

\paragraph{Privacy Is Not in a Vacuum: Supporting Tradeoff Management}

Many Privacy-Enhancing Technologies put optimizing privacy at the center of the aim, whereas this is rarely the case in real life human decision making. In practice, privacy decisions often conflict with factors such as utility, convenience, and monetary cost.
Autonomous agents further complicate the problem by introducing tension between personalization, privacy, and autonomy~\citep{zhangautonomy2025}.
However, humans are susceptible to manipulation, and perceived versus actual protection may diverge~\citep{zhou2025rescriber}.
Therefore, more automated or semi-automated approaches to quantifying and optimizing privacy-utility tradeoffs, coupled with awareness mechanisms and balanced human control and agent autonomy, are needed to achieve an alignment with human interests.
For example, PAPILLON~\citep{siyan2024papillon} demonstrates how local-remote model delegation can balance response quality with reduced privacy leakage.
Beyond privacy-utility balancing, data minimization offers another strategy: it prioritizes utility (or other objectives) while ensuring the least amount of sensitive information is disclosed. Recent work has explored data minimization both as a user-facing inpu privacy control~\citep{zhou2025rescriber, zhou2026autominimization} and as a guiding principle for calibrating disclosure in agent behavior~\citep{zharmagambetov2025agentdam}.

\paragraph{Observability Challenges in Understanding Real-world Impact}

Although our analysis uncovers a small body of work auditing adversarial capabilities in controlled settings, we argue that this does not replace the need to audit adversarial usage in the wild, which presents significant challenges. \citet{vekaria2025big} (USENIX Security 2025) conduct a large-scale audit of generative AI assistants, focusing on how personalization, profiling, and tracking practices may covertly misuse user data.
Large-scale measurement efforts (e.g., GPTracker~\citep{shen2025gptracker}) show promise, but observational data is inherently incomplete and biased: people may deliberately conceal their use of AI~\citep{zhang2025secret}, or avoid disclosure in professional settings where AI use can invite stigma or delegitimization~\citep{sarkar2025ai}. 

Beyond raw measurement, there is also the challenge of communicating findings across disciplinary boundaries. In this paper, we contribute by systematically mapping attacks and defense techniques to observed real-world incidents, exposing gaps where pressing risks remain unaddressed by existing technologies and research agendas. We advocate for more measurement efforts, conducted periodically and continuously.

\subsection{Policy and Governance}

We want to highlight that technical and socio-technical approaches alone cannot completely address the five types of personal data incidents in LLMs that we have identified.
For example, the asymmetric power relationship between LLM provider companies and users, users' lack of AI and privacy literacy, as well as the complex tradeoffs between privacy and other factors such as usability, utility, and monetary values, can easily give rise to manipulative design practices and dark patterns, as illustrated in many of the incidents we discussed.
As autonomous LLM agents become more widely adopted and act as ``netizens'' on behalf of human users, the characterization of manipulative behaviors and the definition of dark patterns may need to be updated to account for the unique vulnerabilities of LLMs.
In particular, such updates should be considered in light of laws such as the FTC Act Section 5, which prohibits unfair or deceptive acts or practices. Extending these protections to LLM-mediated interactions would help ensure that deceptive design choices or manipulative outputs generated by LLMs are evaluated with the same seriousness as traditional dark patterns affecting consumers~\citep{tang2025dark}.

The adversarial use of LLMs, as illustrated in Indirect Attribute Inference and Direct Attribute Aggregation, requires significant support from regulatory and policy perspectives and raises new challenges. On the one hand, such adversarial uses can invade individuals' privacy and are difficult to detect and disable, particularly when they prioritize stealthiness and turn to decentralization or local inferences.
However, they also prompt broader privacy debates with respect to accessing and retaining user chat data for abuse monitoring purposes, as exemplified by the New York Times vs. OpenAI case.

\section{Conclusion}

The privacy challenges posed by LLM systems extend far beyond the narrow technical problem of training data memorization. From deceptive data collection to inference attacks, from context aggregation to autonomous agent risks, the privacy landscape demands comprehensive, interdisciplinary solutions. We argue that the research community must expand its focus beyond memorization to address these pressing, real-world privacy threats. Only through this broader lens can we develop LLM systems that respect user privacy while delivering on their transformative potential.

The path forward requires collaboration between technologists, designers, policymakers, ethicists, and affected communities. As LLMs become increasingly integrated into daily life, the urgency of addressing these ``thousand other things'' beyond memorization cannot be overstated. The privacy iceberg runs deep, and we must map its full extent before it's too late.

\clearpage

\bibliography{iclr2025_conference}
\bibliographystyle{iclr2025_conference}

\appendix

\end{document}

%% file: math_commands.tex
\usepackage{amsmath,amsfonts,bm}

\def\eqref#1{equation~\ref{#1}}
\def\1{\bm{1}}

\DeclareMathAlphabet{\mathsfit}{\encodingdefault}{\sfdefault}{m}{sl}
\SetMathAlphabet{\mathsfit}{bold}{\encodingdefault}{\sfdefault}{bx}{n}

%% file: tables/table_overview.tex
\begin{table}
\centering
\caption{\small Taxonomy of personal data incidents in LLMs: We divide the incidents into five categories, over three different data types: \emoji{speechballoon}~User interactions~(\S\ref{sec:interactions}), \emoji{scroll}~Retrieved documents~(\S\ref{sec:rag}), and \emoji{globewithmeridians}~Publicly available data~(\S\ref{sec:public}). Victim is the person or entity whose data is revealed and data viewer is the entity that gains access to this revealed data, maliciously or by accident.}
\label{tab:taxonomy}
\resizebox{\textwidth}{!}{%
\begin{tabular}{c p{3.8cm}p{3.4cm}p{2.8cm}p{3.9cm}p{3.1cm}}
\toprule
\textbf{Section} & \textbf{Incident Type} & \textbf{Target Data} & \textbf{Victim} & \textbf{Data Viewer} & \textbf{Model Role} \\
\midrule
\S\ref{sec:regurg} &
{Training Data Leakage via Regurgitation} & 
\celllines{\emoji{speechballoon}~User~interactions\\ \emoji{globewithmeridians} Public data }&
\celllines{
\emoji{womantechnologist} User
\\ \emoji{person}Bystander  w/ \\public data   } & 
\celllines{
   \emoji{laptop} Innocent user \\
   \emoji{mansupervillain} Malicious user \\
   \emoji{eyes} Innocent bystander
} & 
Model as data-store \\
\midrule
\S\ref{sec:direct-leak} &
Direct Chat Leakage via Uninformed Consent or Compromised Provider & 
\celllines{\emoji{speechballoon}~User~interactions\\(Full transcript)} & 
\emoji{womantechnologist} User& 
\celllines{
\emoji{eyes} Innocent bystander \\ \emoji{balancescale} Legal proceedings \\ \emoji{mansupervillain} Malicious 3rd party} & 
Model not directly involved \\
\midrule
\S\ref{sec:indirect-leak} &
Indirect Chat and Context Leakage via Input-Output Flow & 
\celllines{ \emoji{speechballoon}~User~interactions\\ \emoji{scroll} Retrieved documents \\ or  data via API } & 
\emoji{womantechnologist} User & 
\celllines{\emoji{mansupervillain} Malicious 3rd party\\ \emoji{eyes} Innocent bystander} & 
Model as autonomous agent \\
\midrule
\S\ref{sec:micrs} &
Indirect   Attribute Inference& 
  \emoji{globewithmeridians}Available data fed to LLM to infer age, location, etc.& 
\emoji{person}  Bystander & 
\emoji{mansupervillain} Malicious user & 
Model as inference engine \\
\midrule
\S\ref{sec:teles} &
Direct Attribute Aggregation & 
\celllines{\emoji{globewithmeridians} Public data: \\finding exact attributes \\via deep research}& 
\celllines{\emoji{person}  Bystander  w/ \\ public data} & 
\emoji{mansupervillain} Malicious user & 
Model as search engine \\
\bottomrule
\end{tabular}
}%
\end{table}